\documentclass[twocolumn,showpacs,preprintnumbers,amsmath,amssymb]{revtex4}

\usepackage[dvips]{graphicx}
\usepackage{dcolumn}
\usepackage{bm}

\begin{document}

\preprint{APS/123-QED}

\title{Critical phenomena of the Majority voter model in a three dimensional cubic lattice}

\author{Ana L.\ Acu\~na-Lara}
\email{ana_lara@fisica.ugto.mx}
\affiliation{%
Departamento de Ingenier\'ia F\'isica,\ Divisi\'on de Ciencias e Ingenier\'ias,\\
Campus Le\'on de la Universidad de Guanajuato%
}
\author{Francisco Sastre}%
 \email{sastre@fisica.ugto.mx}
\affiliation{%
Departamento de Ingenier\'ia F\'isica,\ Divisi\'on de Ciencias e Ingenier\'ias,\\
Campus Le\'on de la Universidad de Guanajuato%
}%

\date{\today}

\begin{abstract}
In this work we investigate the critical behavior of the three dimensional 
simple-cubic Majority voter model. Using numerical simulations and a combination 
of two different cumulants we evaluated the critical point with a higher
accuracy than the previous numerical result found by Yang {\em et al.}~
[J.-S. Yang, I.-M. Kim and W. Kwak, Phys.\ Rev.\ E \textbf{77}, 051122 (2008)].
Using standard Finite Size Scaling theory and scaling corrections we find that 
the critical exponents $\nu, \gamma$ and $\beta$ are the same as those of the 
three dimensional Ising model.
\end{abstract}

                             
\maketitle

\section{INTRODUCTION}
The Majority Voter (MV) model is one of the simplest non-equilibrium 
models that present a second order phase transition, and its critical 
exponents for a two-dimensional square lattices are the same as those 
of the Ising model~\cite{Oliveira92,Kwak2007}. Those results confirm 
the conjecture that non equilibrium models with up-down symmetry and 
spin flip dynamics fall within the universality class of the equilibrium 
Ising model~\cite{Grinstein}. However, other numerical results suggest
that the MV on non-regular lattices does not belong to the Ising 
universality class~\cite{Campos2003,Lima2005,Lima2006,Wu2010}.

In a recent work Yang {\em et al.}~\cite{Yang2008} carried out Monte 
Carlo simulations for the MV model on regular lattices from three to seven 
dimensions, and they found that the critical exponents differ from the 
Ising ones below $d=6$. Based on this results they suggest that the 
upper critical dimension for the MV model is 6 instead of 4. This is 
a really important result that requires verification, since it implies
the existence of a new universality class for Ising-like spin systems 
in regular lattices. Another result that raises some concerns relates 
to the Rushbrooke and Josephson hyperscaling relation, which is not 
satisfied with the reported exponents for the three-dimensional case. 
In fact,~\cite{Yang2008} obtains 
$(2 \beta+\gamma)/\nu = 3.30(1)$, and this implies that there is an 
effective non-integer dimension, a fact that can not be overlooked.

The aim of this work is to study the critical phenomena of the 
three-dimensional MV model on a cubic lattice, using Monte
Carlo simulations and taking into account the effects of the 
leading correction to scaling in the evaluation of the critical 
exponents. In this way we expect to clarify the universality 
class for this model.

\section{Model}
\label{model}

The MV model in two-dimensional lattices belongs to a family 
of two-dimensional non-equilibrium kinetic spin models introduced 
some time ago in~\cite{Oliveira93}, and defined by the following 
evolution rules: During an elementary timestep, an Ising-like spin 
$\sigma_i = \pm 1$ on a square lattice is randomly picked up, and 
flipped with a probability given by
    \begin{equation}
    p(x)= \left\{ \begin{array}{ccc}
                \frac{1}{2}(1+x) & \mbox{if} & H_{i}\cdot\sigma_{i}< 0 \\
                \frac{1}{2}      & \mbox{if} & H_{i}=0    \\
                \frac{1}{2}(1-x) & \mbox{if} & H_{i}\cdot\sigma_{i}> 0
              \end{array}
    \right..%
    \label{transition}
    \end{equation}
Here $H_i$ is the local field produced by the four nearest neighbors 
to the $i$-th spin and $x$ is the control parameter (coupling). 
The system presents a continuous phase transition from an disordered state (paramagnetic-like phase)
to an ordered one (ferromagnetic-like phase) as $x$ is increased, the reported value for the critical point 
$x_c$ in two-dimensional lattices is $0.8500(4)$~\cite{Kwak2007}. 
This evolution rule can be used also in a three-dimensional 
cubic lattice, since the updating prescription depends 
only on the sign of the local field $H_i$. This 
definition is fully equivalent to the used in~\cite{Yang2008} with 
$x=\tanh(1/T)$.

The instantaneous order parameter $m_t$ is defined as an spin average over 
all lattice sites in each Monte Carlo Time Step (MCTS)
\begin{equation}
m_t =  \frac{1}{N} \sum_i \sigma_i,
\end{equation}
where $N=L^3$ is the total number of lattice sites and $L$ is the linear 
dimension. From here we can evaluate the moments of the order parameter 
as time averages 
    \begin{equation}
    \langle m^k\rangle =  \frac{1}{T-\tau} \sum_{t=\tau}^T |m_t|^k,
    \label{orderparameter}
    \end{equation}
where $\tau$ is the transient time and $T-\tau$ is the running time. 
The Susceptibility is given by
     \begin{equation}
     \chi = N x \{\langle m^2\rangle-\langle m\rangle^2\}.
     \label{susceptibility}
     \end{equation}
We will use two different cumulants in order to locate the critical 
point, the fourth order cumulant~\cite{binder} (commonly known as 
Binder Cumulant)
    \begin{equation}
    U^4=1-\frac{\langle m^4\rangle}{3\langle m^2\rangle^2},
    \label{cumulant4}
    \end{equation}
which is  often used to locate the critical point, and the second 
order cumulant~\cite{Deutsch92,Gabriel}
    \begin{equation}
    U^2=1-\frac{2\langle m^2\rangle}{\pi\langle m\rangle^2}.
    \label{cumulant2}
    \end{equation}
In the next section we will explain how both cumulants can be 
combined to improve the estimation of the critical point.

\section{Finite Size Scaling}

Finite Size Scaling theory establishes that it is possible 
to know the critical properties of an infinite system, in 
particular, its critical exponents and amplitude ratios, using a 
set of finite systems of increasing linear sizes that obey
the same microscopic dynamics of the infinite one. We assume 
that, even though we are working with a non-equilibrium model, 
the same scaling forms used in the equilibrium models can be 
applied. So, we start with the following fact: For an 
equilibrium finite systems of linear size $L$, with couplings 
close to those of the critical point that appears at its 
$L \to \infty$ limit, the free energy density is given 
by the scaling ansatz
    \begin{equation}
    F(x,h,L)\approx  L^{-(2-\alpha)/\nu}f^0(\epsilon 
         L^{1/\nu},hL^{(\beta+\gamma)/\nu}),
    \label{scaling}
    \end{equation}
where $\epsilon=(x-x_c)$, $x_c$ is the critical point 
for the infinite system, $f^0$ is a universal function 
and $h$ is the symmetry-breaking (magnetic) field. The 
parameters $\alpha$, $\beta$, $\gamma$ and $\nu$ are the 
critical exponents for the infinite system. 
From~(\ref{scaling}) the scaling forms for the thermodynamic 
observables can be obtained, with $h=0$, as
     \begin{equation}
     \begin{array}{lcl}
     m &= &L^{-\beta/\nu}\hat{M}(\epsilon L^{1/\nu}),\\
     \chi &=  &L^{\gamma/\nu}\hat{\chi}(\epsilon L^{1/\nu}),\\ 
     U^p &=  &\hat{U}^p(\epsilon L^{1/\nu}).
     \end{array}\label{scaling2}
     \end{equation}
In principle the scaling relations~(\ref{scaling2}) can 
be used to evaluate the critical exponents when $L$ is 
sufficiently large. For smaller systems scaling corrections, 
presents as power law corrections, must be taken into account. 
Considering one leading 
correction exponent $\omega$ the scaling relations behave as
    \begin{eqnarray}
    m(\epsilon,L) &\approx&  L^{-\beta/\nu}(\hat{M}(\epsilon L^{1/\nu})+
          L^{-\omega} \hat{\hat{M}}(\epsilon L)),\\
    \chi(\epsilon,L) &\approx&  L^{-\gamma/\nu}(\hat{\chi}(\epsilon 
          L^{1/\nu})+L^{-\omega} \hat{\hat{\chi}}(\epsilon L)),\\
    U^p(\epsilon,L) &\approx& \hat{U}^p(\epsilon L^{1/\nu})+
          L^{-\omega} \hat{\hat{U}}(\epsilon L).\label{scaling3}
    \end{eqnarray}
Setting $\epsilon=0$ we obtain the following set of equations that 
allow us to evaluate the critical exponents for small lattice sizes:
    \begin{eqnarray}
    m(L) &\propto& L^{-\beta/\nu}(1+ a L^{-\omega}), \label{beta}\\
    \chi(L) &\propto& L^{\gamma/\nu}(1+ b L^{-\omega}), \label{gamma}
    \end{eqnarray}
and
    \begin{equation}
    \frac{\partial U^p}{\partial x}\Bigl|_{x=x_c} 
        \propto L^{1/\nu}(1+c_p L^{-\omega}).
    \label{nu}
    \end{equation}
The parameters $a,~b$ and $c_p$ are non-universal constants. 
In order to use equations~(\ref{beta})$-$(\ref{nu}) we need to 
evaluate with good accuracy the critical point $x_c$. In this work 
we are using an approach suggested by P\'erez~\cite{Gabriel} that 
is based on observing the differences between the crossing points 
in $U^2$ for different values of $L$, with respect to the corresponding
crossings evaluated for $U^4$. The method take into account the 
correction-to-scaling effects on the crossing points. First it 
is necessary to expand Eq.~(\ref{scaling3}) around $\epsilon=0$ 
(that is, around the critical coupling), to obtain
    \begin{equation}
    U^p\approx U^p_\infty+\bar{U}^p \epsilon L^{1/\nu} + 
        \bar{\bar{U}}^p L^{-\omega}+\mathcal{O}(\epsilon^2,\epsilon L^{-\omega}).
    \label{crossing1}
    \end{equation}
Here the $U^p_\infty$ are universal quantities, but $\bar{U}^p$ and 
$\bar{\bar{U}}^p$ are non-universal. The value of $\epsilon$ where 
the cumulant curves $U^p$ for two different linear sizes $L_i$ and 
$L_j$ intercept is denoted as $\epsilon^p_{i,j}$. At this crossing 
point the next relation must be satisfied
     \begin{equation}
     L_i^{1/\nu}\epsilon^p_{ij}+B^p L_i^{-\omega} =
     L_j^{1/\nu}\epsilon^p_{ij}+B^p L_j^{-\omega}.
     \end{equation}
Here $B^p=\bar{\bar{U}}^p/\bar{U}^p$. We next get the relations
    \begin{equation}
    \epsilon^2_{ij}=B^2\frac{L_j^{-\omega}-L_i^{-\omega}}{L_j^{1/\nu}-L_i^{1/\nu}}
    \end{equation}
and \cite{binder}
    \begin{equation}
    \epsilon^4_{ij}=B^4\frac{L_j^{-\omega}-L_i^{-\omega}}{L_j^{1/\nu}-L_i^{1/\nu}},
    \label{crossing2}
    \end{equation}
that can be used to evaluate the critical point. However, 
in order to avoid nonlinear fittings we need to get rid of the 
dependence of these expressions on $\nu$ and $\omega$.
The presence of different $B^p$ coefficients allow us to do this,
using the combination of the last two equations to get
    \begin{equation}
    \frac{(x^{p}_{ij}+x^{q}_{ij})}{2}=x_{c}-A_{pq}(x_{ij}^{p}-x_{ij}^{q}),
    \label{linealcrit}
    \end{equation}
Here $A_{pq}=(B^p+B^q)/2(B^p-B^q)$ and $x^p_{i,j}=\epsilon^p_{i,j}+x_c $. 
Eq.~(\ref{linealcrit}) is a linear equation that makes no reference to
$\nu$ or $\omega$, and requires as inputs only the numerically
measurable crossing couplings $x_{i,j}^p$. The intercept with the ordinate
provides an improved estimate of the critical coupling.

\section{Results}
Our simulations where carried out on a three-dimensional simple cubic 
lattice with periodic boundary conditions, and the linear sizes used 
were $L=12,~14,~16,~20,~24$ and $28$. Starting with a random configuration 
of spins the system evolves following the dynamic rule explained in 
section~\ref{model}. Even though the MV model does not satisfy the
detailed balance condition, it has stationary probability distribution 
functions. The stationary state is reached after a transient time, which 
in this work varied from $5\times10^4$ MCTS for $L=12$ to $1.5\times10^5$ 
MCTS for $L=28$. Averages of the observables were taken over 
$4\times 10^5$ MCTS for $L=12$ and up to $1.5\times 10^6$ MCTS for $L=28$. 
Additionally, for each value of $x$ and $L$ we performed up to 160 independent 
runs in order to improve the statistics. 

\begin{figure}
     \includegraphics[width=8.5cm,clip]{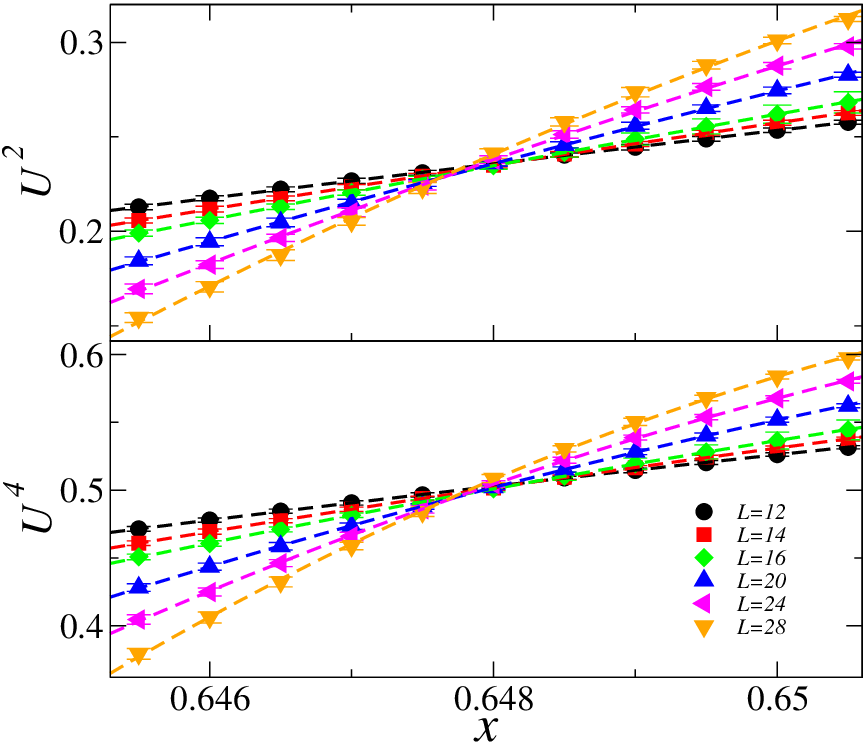}
     \caption{\label{cumulants} (color online) Curve crossings for 
     the second (upper graph) and fourth (lower graph) order cumulants 
     for various lattice sizes as function of the control parameter $x$.
     The crossing region is the same in both 
     cases. Dashed lines are third order polynomial fittings.}
     \end{figure}

\begin{figure}
    \includegraphics[width=7.5cm,clip]{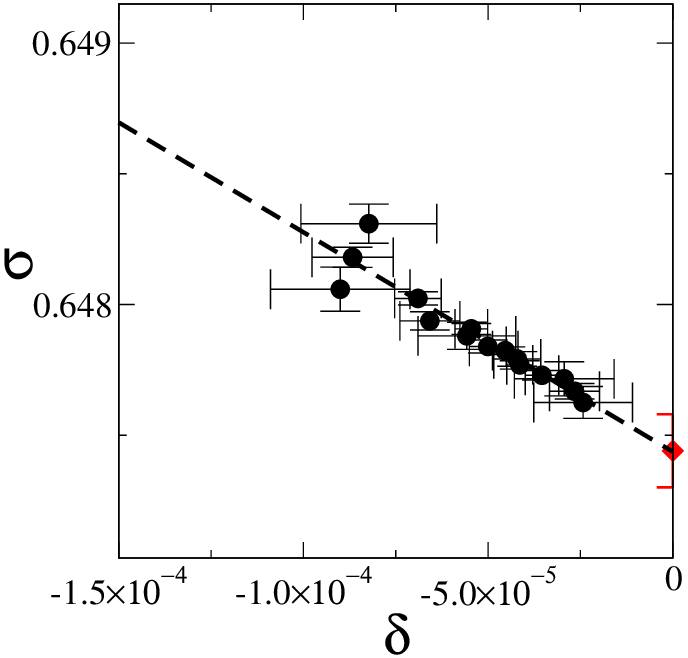}
    \caption{\label{criticalpoint} (color online) Evaluation of the critical 
    point using the quantities $\sigma=(x^{4}_{ij}+x^{2}_{ij})/2$ and
    $\delta= x_{ij}^{4}-x_{ij}^{2}$, where $x^q_{ij}$ is the 
    crossing point between cumulants $U^q$ for linear sizes $L_i$ and $L_j$. 
    The dashed line is the linear fit and the 
    red diamond shows the critical point.}
    \end{figure}

Fig.~\ref{cumulants} shows the cumulants curves for the different 
linear sizes around the critical point. We used third order 
polynomial fitting in each curve to obtain the crossing points between 
each pair of curves.
The estimation of the
critical point is shown in Figure~\ref{criticalpoint}, where we use the 
notation $\delta=x_{ij}^4-x_{ij}^2$ and $\sigma=(x_{ij}^4+x_{ij}^2)/2$. 
The linear fit gives an estimated for the critical point of 
$x_c=0.64744(14)$ that is in a good agreement with the reported by 
Yang {\em et al.} of $x_c=0.646(2)$. Our result improve the previous 
one by one order of magnitude. 
The leading correction exponent $\omega$ and the universal quantities 
$U^2_\infty$ and $U^2_\infty$ can be obtained using~(\ref{crossing1}) 
at the critical point, using a non-linear curve fitting. Our computed 
values are $\omega=0.6(2)$, $U^2_\infty= 0.207(8)$ and 
$U^4_\infty= 0.458(12)$.
Our results for the cumulants are in good agreement with the reported values
$U^2 = 0.2108(7)$~\cite{Lundow2010} and 
$U^4 = 0.4656(4)$~\cite{Ballesteros99} 
of the three-dimensional Ising model. 
Our result for the $\omega$ exponent is clearly 
smaller, although within error bar ranges, that previous reported results: 
$\omega=0.86(9)$~\cite{Ballesteros99}, 
$\omega=0.814(18)$~\cite{Guida98} and
$\omega=0.782(5)$~\cite{Pogorelov2008}; 
because of this discrepancy we will evaluate the critical exponents 
using both our $\omega$ value and a fixed value of $\omega=0.8$, in order
to check the validity of our results.

\begin{figure}
    \includegraphics[width=8.5cm,clip]{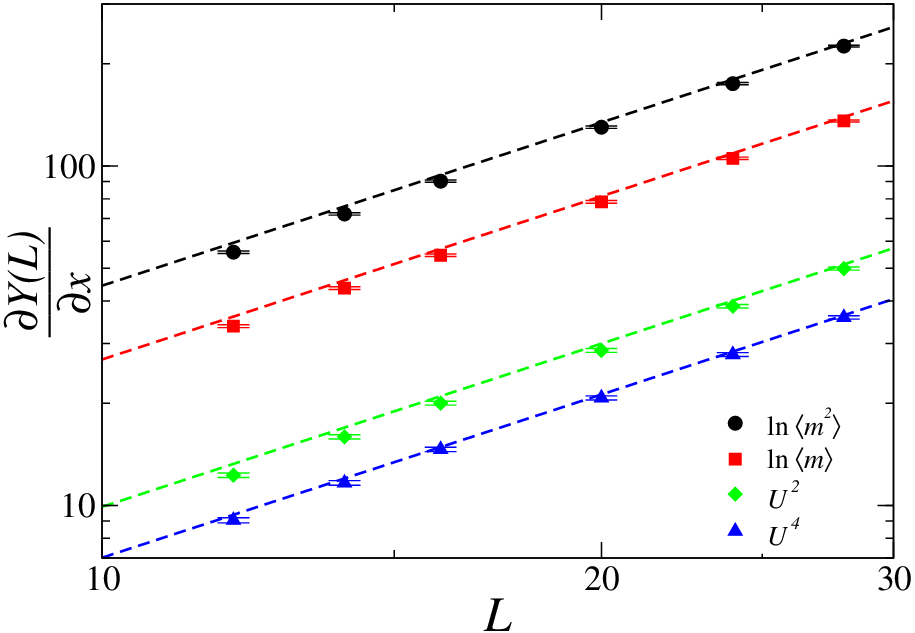}
    \caption{\label{nuexponent} (color online)
    Log-log plot of several derivatives at the critical point as function of the
    linear dimension $L$, 
    from top to bottom:
    $Y(L)=\ln\langle m^2\rangle,~\ln\langle m\rangle$,~$U^2$ 
    and $U^4$.
    The dashed lines show the expected power law behavior in the $L\to\infty$ limit with
    $1/\nu=1.597$.}
    \end{figure}

For the evaluation of the critical exponent $\nu$ we use (\ref{nu}) 
with both cumulants $U^2$ and $U^4$. Additionally we use the logarithmic 
derivative of $\langle m\rangle$ and $\langle m^2 \rangle$, which 
have the same scaling properties of the cumulant slope~\cite{Ferrenberg91}
    \begin{equation}
    \frac{\partial \ln \langle m^n\rangle}{\partial x}\Bigl|_{x=x_c} 
         \propto L^{1/\nu}(1+z_n L^{-\omega}).
    \label{nu2}
    \end{equation}
Here the $z_n$ are non universal constants. In Figure~\ref{nuexponent} 
we are showing the derivatives of the thermodynamic quantities used to 
evaluate $1/\nu$.
The results for $1/\nu$ from the fits with two different values of $\omega$ 
are given in Table~\ref{nuresults}. We observe that the $U^2$ fitting 
give the largest error of all.
    \begin{center}
    \begin{table}
    \caption{\label{nuresults} Estimates for $1/\nu$ obtained from the 
    fitting of Eqs.~(\ref{nu}) and (\ref{nu2})
    at the critical point $x_c=0.64744(14)$.}
    \begin{tabular}{cccc}
    \hline
    \hline
     ~ & ~ & $1/\nu$ & ~ \\
    \hline
    \hline
     ~& $\omega=0.6$ & ~& $\omega=0.8$ \\
    \hline
    $U^2$ & 1.617(51)& ~ & 1.622(43) \\
    $U^4$ & 1.599(12)& ~ & 1.602(10) \\
    $\ln \langle m\rangle$ & 1.594(16)& ~ & 1.603(13) \\
    $\ln \langle m^2\rangle$ & 1.592(20)& ~ & 1.603(17) \\
    \hline
    \hline
    \end{tabular}
    \end{table}
    \end{center}
Combining the results we get $1/\nu=1.597(29)$ or $\nu=0.626(11)$ for 
$\omega=0.6$ and $1/\nu=1.604(24)$ or $\nu=0.623(9)$ for $\omega=0.8$. 
Both results are in good agreement with the results
$\nu=0.6305(5)$,
obtained by field 
theoretical renormalization group method (RG)~\cite{Pogorelov2008}, and
$\nu=0.6302(1)$, obtained by Monte Carlo simulations (MC)~\cite{Campbell2011},
for the three-dimensional Ising model.

\begin{figure}
     \includegraphics[width=8.5cm,clip]{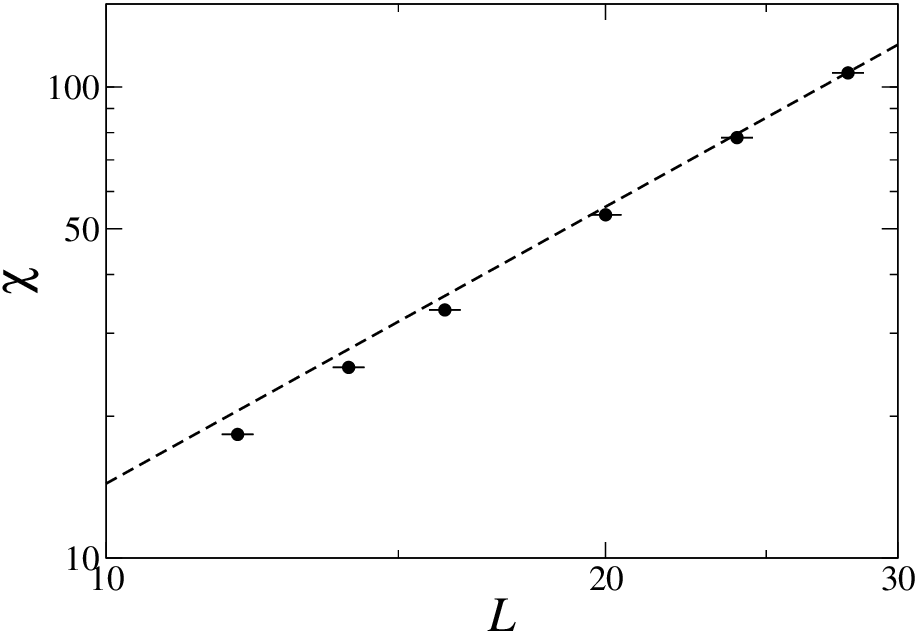}
     \caption{\label{gammaexponent}
     Log-log plot of the susceptibility at the critical point as function of the
     linear dimension $L$.
    The dashed lines show the expected power law behavior in the $L\to\infty$ limit
     with $\gamma/\nu=1.952$.}
\end{figure}

For the critical exponent $\gamma$ we are using Eq.~(\ref{gamma}) to 
fit our data at the critical point (see Figure~\ref{gammaexponent}).
Our results are $\gamma/\nu=1.952(11)$ or $\gamma=1.222(23)$ for 
$\omega=0.6$ and $\gamma/\nu=1.976(9)$ or $\gamma=1.232(19)$ for 
$\omega=0.8$. Again the agreement is acceptable compared with the values
$\gamma=1.2411(6)$ (RG)~\cite{Pogorelov2008} and $\gamma=1.2372(4)$ (MC)~\cite{Campbell2011}.

\begin{figure}
    \includegraphics[width=8.5cm,clip]{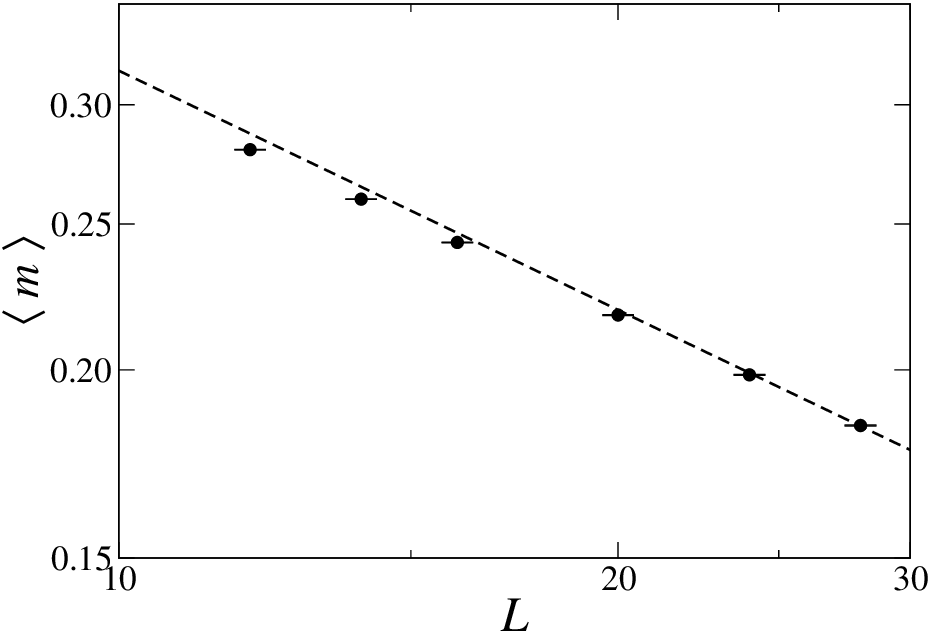}
    \caption{\label{betaexponent}
    Log-log plot of the order parameter at the critical point as function of the
    linear dimension $L$.
    The dashed lines show the expected power law behavior in the $L\to\infty$ limit
    for $\beta/\nu=0.528$.}
    \end{figure}

The fitting for the $\beta$ exponent is shown in Figure~\ref{betaexponent}. 
Our estimates are $\beta/\nu=0.528(54)$ or $\beta=0.331(34)$ for $\omega=0.6$,
 and $\beta/\nu=0.521(45)$ or $\beta=0.325(28)$ for $\omega=0.8$. 
We compare our results with $\beta=0.3253(8)$ (RG) and $\beta=0.3267(2)$, built with the values 
of $\gamma$ and $\nu$ reported in~\cite{Pogorelov2008} and \cite{Campbell2011}, observing that 
all results are in good agreement.

One important point of our results is that in both cases the Rushbrooke 
and Josephson hyperscaling relation is satisfied, we obtain 
$(\gamma+2\beta)/\nu=3.008(55)$ for $\omega=0.6$ and 
$(\gamma+2\beta)/\nu=3.012(46)$ for $\omega=0.8$. In 
Table~\ref{finalresults} we summarize the results for the MV model 
from this work, the values obtained previously by Yang {\em et al.},
and the known values of the three-dimensional Ising model obtained by Monte Carlo Simulations.
      \begin{center}
      \begin{table}
      \caption{\label{finalresults} Critical values for the MV and the Ising models 
      in three-dimensional lattices.}
      \begin{tabular}{c|cccc}
      \hline
      \hline
       ~ & This work $(\omega=0.6)$ & Yang {\em et al.}~\cite{Yang2008} & Ising  \\
      \hline
      $U^2_\infty$ & $0.207(8)$ & $-$ & $0.2108(7)$~\cite{Lundow2010} \\
      $U^4_\infty$ & $0.458(12)$ & $-$ & $0.4656(4)$~\cite{Ballesteros99} \\
      $\nu$ & $0.626(11)$ & $ 0.63(1)$  & $0.6302(1)$~\cite{Campbell2011} \\
      $\gamma$ & $1.222(23)$ & $ 1.32(3) $ & $1.2372(4)$~\cite{Campbell2011} \\
      $\beta$ & $0.331(34)$ & $ 0.38(1) $ & $0.3267(2)$~\cite{Campbell2011} \\
      $(\gamma + 2\beta)/\nu$  & $3.008(55)$ & $ 3.30(1) $ & $3.0000(12)$~\cite{Campbell2011} \\
      \hline
      \hline
      \end{tabular}
      \end{table}
      \end{center}

\section{Conclusions}
The MV model on three-dimensional simple cubic lattices belongs 
to Ising model universality class. Our simulations prove that 
the set of critical exponents for both models are consistent 
when corrections to scaling are included. The incertitude in 
the value of $\omega$ indicates that it is necessary to increase
the simulation data in order to improve the accuracy of the 
leading correction exponent. However, we have shown that the results 
are not considerably affected by the choice between 
$\omega=0.6$ or $\omega=0.8$.
In this case the conjecture of Grinstein
{\em et al.}~\cite{Grinstein} is satisfied, however,
we believe that additional numerical simulations
in the fourth-dimensional case should be made
in the future in order to corroborate 
the conclusion made by Yang {\em et al.}~\cite{Yang2008} 
about the critical dimension for the MV model. 
One open interesting
topic is whether or not the dynamical critical phenomena 
of the MV model is the same that in the Ising model.
There are some works that indicate that it is the case for 
two-dimensional systems~\cite{Mendes98,Tome98,Sastre03},
but results for larger dimensionalities are non-existent.

\section{Acknowledgments}
We wish to thank to G.\ P\'erez for helpful comments and suggestions. 
A.\ L.\ Acu\~na-Lara thanks Conacyt (M\'exico) for
fellowship support. This work was supported by Conacyt (M\'exico)
through Grant No. 61418/2007.

\end{document}